\documentstyle[graphicx,aps,multicol]{revtex}  

\begin{document}  

\title{Heat capacity and pairing transition in nuclei}  \author{M.~Guttormsen\footnote{Electronic address: magne.guttormsen@fys.uio.no}, M.~Hjorth-Jensen, E.~Melby, J.~Rekstad, A.~Schiller\footnote{Present address: Lawrence Livermore National Laboratory, L-414, 7000 East Avenue, Livermore CA-94551}, and S.~Siem} \address{Department of Physics, University of Oslo, P.O.Box 1048 Blindern, N-0316 Oslo, Norway}  

\maketitle  

\begin{abstract} 
A simple model based on the canonical-ensemble theory is outlined for hot nuclei. The properties of the model are discussed with respect to the Fermi gas model and the breaking of Cooper pairs. The model describes well the experimental level density of deformed nuclei in various mass regions. The origin of the so-called S-shape of the heat capacity curve $C_V(T)$ is discussed. 
\end{abstract}  


\begin{multicols}{2}   

\section{Introduction}  

Nuclear structure at low excitation energy depends critically on the presence of Cooper pairs. Thermal and rotational breaking of these $J=0$ nucleon pairs gives abrupt structural changes, such as increased level density and rotational-spin alignments. These critical phenomena were early addressed in several theoretical papers \cite{Sano,Tanabe,Goodman}.

A very exciting feature is the gradual reduction of pair correlations as function of temperature. Recently, Schiller et al.~\cite{Schi1} reported the experimental critical temperature $T_c$ for the pairing transition. The findings were based on using the canonical heat capacity as thermometer. An S-shaped heat capacity as function of temperature was observed in the $^{161,162}$Dy and $^{171,172}$Yb isotopes. Around $T_c \sim 0.5$~MeV, a local maximum in the heat capacity signals the breaking of Cooper pairs and quenching of the pair correlations. This property has also been observed in the $^{166,167}$Er nuclei \cite{Melb1}. 

Similar fingerprints have been obtained in various calculations. Finite-temperature Hartree Fock Bogoliubov calculations~\cite{Egido} for $^{164}$Er give almost identical S-shape as observed for $^{166}$Er. In relativistic Hartree Fock-BCS calculations~\cite{Agrawal} the proton and neutron pairing gaps are seen to vanish around $T\sim 0.4 - 0.5$ MeV for $^{166,170}$Er. Furthermore, in Shell Model Monte Carlo simulations (SMMC) \cite{Rombouts,Liu}, the heat capacities for iron isotopes show a pairing transition around temperatures of $0.7$ MeV.    

The thermal breaking of a Cooper pair results in a tenfold increase in number of available energy levels. In this process particles are thermally scattered on available single particle states, giving rise to increased entropy. Recently\cite{Gutt1,Gutt2}, it was shown that each thermal particle carries an entropy of $\sim 1.7$, a feature which is valid for midshell nuclei with mass number $A>40$.  

The present work aims to present a simple model for hot nuclei that includes the main features found experimentally. In Sec.~II the model is described within the canonical-ensemble theory, and in Sec.~III some relevant model properties are discussed. The results are compared with recent theoretical and experimental data on various deformed nuclei in Sec.~IV. Concluding remarks are given in Sec.~V.   

\section{Model}  

The total partition function $Z$ is described in the canonical ensemble, where thermal particle excitations, rotations, and vibrations are treated adiabatically according to 
\begin{equation} 
Z(T) \approx Z_{\rm par}Z_{ \rm rot}Z_{\rm vib}. 
\end{equation} 
Thermodynamical quantities such as the entropy $S$, the average excitation energy $\langle E\rangle$ and the heat capacity $C_V$ can then be calculated from the Helmholtz free energy: 
\begin{equation} 
F(T) = -T\ln Z(T), \label{eq:FT} 
\end{equation}
by
\begin{eqnarray}
S(T)&=&-\left(\frac{\partial F}{\partial T}\right)_V, \label{eq:ST} \\
 \langle E(T)\rangle&=&F+TS, \label{eq:ET} \\
 C_V(T)&=&\left(\frac{\partial\langle E\rangle}{\partial T}\right)_V, \label{eq:CV} 
\end{eqnarray}  
where the Boltzmann constant is set to unity ($k_B=1$) and $T$ is measured in units of MeV.

The theoretical basis for the particle partition function $Z_{\rm par}$ was earlier presented in Ref.~\cite{Gutt2}, however, for completeness we make a short outline here. The particle excitations are described by spin $1/2$ fermions scattered into doubly degenerated single-particle levels with equal energy spacing $\epsilon$. There are infinitely many single-particle levels with one set of levels for protons and one set for neutrons. The partition function for one spin-up fermion is given by 
\begin{equation} 
z_1^{\uparrow}=\sum_{i=0}^{\infty}\exp(-i\epsilon/T) =\frac{1}{1-\exp(-\epsilon/T)}. 
\end{equation} 
When two identical spin-up fermions are placed into the same single-particle level scheme, the summing is restricted due to the Pauli principle by 
\begin{eqnarray}
 z_2^{\uparrow}&=&\sum_{i=0}^{\infty}\left[\exp(-i\epsilon/T) \sum_{j=i+1}^{\infty}\exp(-j\epsilon/T)\right]\nonumber\\ &=&\frac{1}{1-\exp(-2\epsilon/T)}\frac{\exp(-\epsilon/T)}{1-\exp(-\epsilon/T)} \nonumber\\ &=&z_1^{\uparrow}\frac{\exp(-\epsilon/T)}{1-\exp(-2\epsilon/T)}. 
\end{eqnarray} 
Generally, for $n$ identical spin-up fermions the partition function is 
\begin{equation} 
z_{n}^{\uparrow}=z_{n-1}^{\uparrow}\frac{\exp(-(n-1)\epsilon/T)} {1-\exp(-n\epsilon/ T)}. 
\end{equation}  

We now introduce spin up {\em and} down for the fermions, and evaluate the corresponding partition function $z_n$. For one fermion with spin up {\em or} down we obtain a degeneration of two, since $z_1^{\uparrow}=z_1^{\downarrow}$: 
\begin{equation}
 z_1=z_1^{\uparrow}+z_1^{\downarrow}=2z_1^{\uparrow}. 
\end{equation} 
If two fermions occupy the same single-particle level scheme, the partition function becomes more complicated. With the spins of the two fermions antiparallel ($m=m_1+m_2=0$), no Pauli blocking is present and 
\begin{equation} 
z_2(m=0)=z_1^{\uparrow}z_1^{\downarrow}=\left(z_1^{\uparrow}\right)^2. 
\end{equation} 
For the contribution with parallel spin, the Pauli principle gives 
\begin{equation}
 z_2(m=\pm 1)=z_2^{\uparrow}+z_2^{\downarrow}=2z_2^{\uparrow}. 
\end{equation} 
It is now straightforward to evaluate $z_n$ for $n$ fermions, allowing spin up and down. The five lowest partition functions are: 
\begin{eqnarray} z_1&=&2z_1^{\uparrow}, \nonumber \\
 z_2&=&2z_2^{\uparrow}+\left(z_1^{\uparrow}\right)^2, \nonumber \\ z_3&=&2z_3^{\uparrow}+2z_2^{\uparrow}z_1^{\uparrow}, \nonumber \\ z_4&=&2z_4^{\uparrow}+2z_3^{\uparrow}z_1^{\uparrow}+ \left(z_2^{\uparrow}\right)^2,\nonumber \\ z_5&=&2z_5^{\uparrow}+2z_4^{\uparrow}z_1^{\uparrow}+ 2z_3^{\uparrow}z_2^{\uparrow}. 
\end{eqnarray} 

The partition function for nuclei with protons and neutrons may now be constructed from $z_n$. In, e.g., the three-particle case of an odd-proton system the $\pi\nu^2$ and the $\pi^3$ partition functions have to be included. For an even-even two-particle system a $\pi$ or a $\nu$ pair can be broken, and so on. The lowest partition functions for odd-mass ($Z_{1,3,5}$) and even-even systems ($Z_{2,4}$) become: 
\begin{eqnarray} 
Z_1& = &z_1, \nonumber \\
  Z_2& = &2z_2, \nonumber \\
 Z_3& = &z_3+z_1z_2, \nonumber \\
 Z_4& = &2z_4+(z_2)^2, \nonumber \\
 Z_5& = &z_5+z_3z_2+z_1z_4. \label{eq:Zeoee} 
\end{eqnarray}  

The partition functions for the odd-odd system can be constructed in a similar manner. The two-particle partition function is $(z_1)^2$, giving an entropy of twice the single particle entropy of the odd-mass system. Higher numbers of particles yield the following partition functions for the odd-odd case: 
\begin{eqnarray}
 \tilde{Z}_2&=&(z_1)^2, \nonumber \\
 \tilde{Z}_4&=&2z_3z_1, \nonumber \\
 \tilde{Z}_6&=&2z_5z_1+(z_3)^2. 
\end{eqnarray}  

The creation of particles costs energy. In order to break one Cooper pair, the energy $2\Delta$ is necessary. The pairing gap parameter is given by the empirical formula \cite{Bohr1}
\begin{equation} 
\Delta=12A^{-1/2} {\rm MeV}. \label{eq:gap} 
\end{equation} 
The total particle partition functions of even-even (ee), odd (odd) and odd-odd (oo) nuclei can then be expressed as 
\begin{eqnarray}
 Z_{\rm par}^{\mathrm{ee}}&=&1+Z_2e^{-2\Delta/T}+Z_4e^{-4\Delta/T}+\ldots, \nonumber \\
 Z_{\rm par}^{\mathrm{odd}}&=&Z_1+Z_3e^{-2\Delta/T}+Z_5e^{-4\Delta/T}+\ldots, \nonumber \\
 Z_{\rm par}^{\mathrm{oo}}&=&\tilde{Z}_2+\tilde{Z}_4e^{-2\Delta/T} +\tilde{Z}_6e^{-4\Delta/T}+\ldots \label{eq:zoo}
\end{eqnarray} 
The number of active particles $n$ (i.e. particles not bound in Cooper pairs) can be evaluated by 
\begin{eqnarray} 
n^{\mathrm{ee\ }}&=& (2Z_2e^{-2\Delta/T}+4Z_4e^{-4\Delta/T}+\ldots) /Z_{\rm par}^{\mathrm{ee}}, \nonumber \\ n^{\mathrm{odd }}&=& (Z_1+3Z_3e^{-2\Delta/T}+5Z_5e^{-4\Delta/T}+\ldots) /Z_{\rm par}^{\mathrm{odd}}, \nonumber \\ n^{\mathrm{oo\ }}&=&(2\tilde{Z}_2+4\tilde{Z}_4e^{-2\Delta/T}+6\tilde{Z}_6e^{-4\Delta/T}+\ldots) /Z_{\rm par}^{\mathrm{oo}}. \label{eq:noo} 
\end{eqnarray}  

The rotational partition function is given by 
\begin{equation}
 Z_{\rm rot} = \sum_{I=0,2,4 \ldots}\exp \left[ -A_{\rm rig}I(I+1)/T \right], 
\end{equation} 
where the rotational parameter  
\begin{equation} 
A_{\rm rig}=\hbar ^2 / 2 \theta_{\rm rig} \label{eq:arot} 
\end{equation} 
is expressed by the rigid moment of inertia \cite{Richter} $\theta_{\rm rig} =2/5MR^2 \sim 0.0137A^{5/3}$ $\hbar^2$MeV$^{-1}$, where $M$ and $R$ are the nuclear mass and radius, respectively. The vibrational partition function includes zero and one-phonon states: 
\begin{equation}
 Z_{\rm vib} = \sum_{v=0,1}W_v\exp ( -v \hbar\omega_{\rm vib}/T), 
\end{equation} 
where the multiplicity for the zero-phonon state is $W_0 = 1$. For one-phonon states, we take into accounts three vibrations (e.g. $\beta$, $\gamma$ and octupole vibrations), giving multiplicity $W_1 = 3$. Higher order phonon states are neglected, and the phonons are assumed to carry the same energy quantum $\hbar\omega_{\rm vib}$. In both partition functions we have omitted the spin degeneracy ($2I+1$), since we are dealing with levels and not states. 

\section{Model properties}  

\begin{figure} 
\includegraphics[totalheight=17.5cm,angle=0,bb=0 80 350 730]{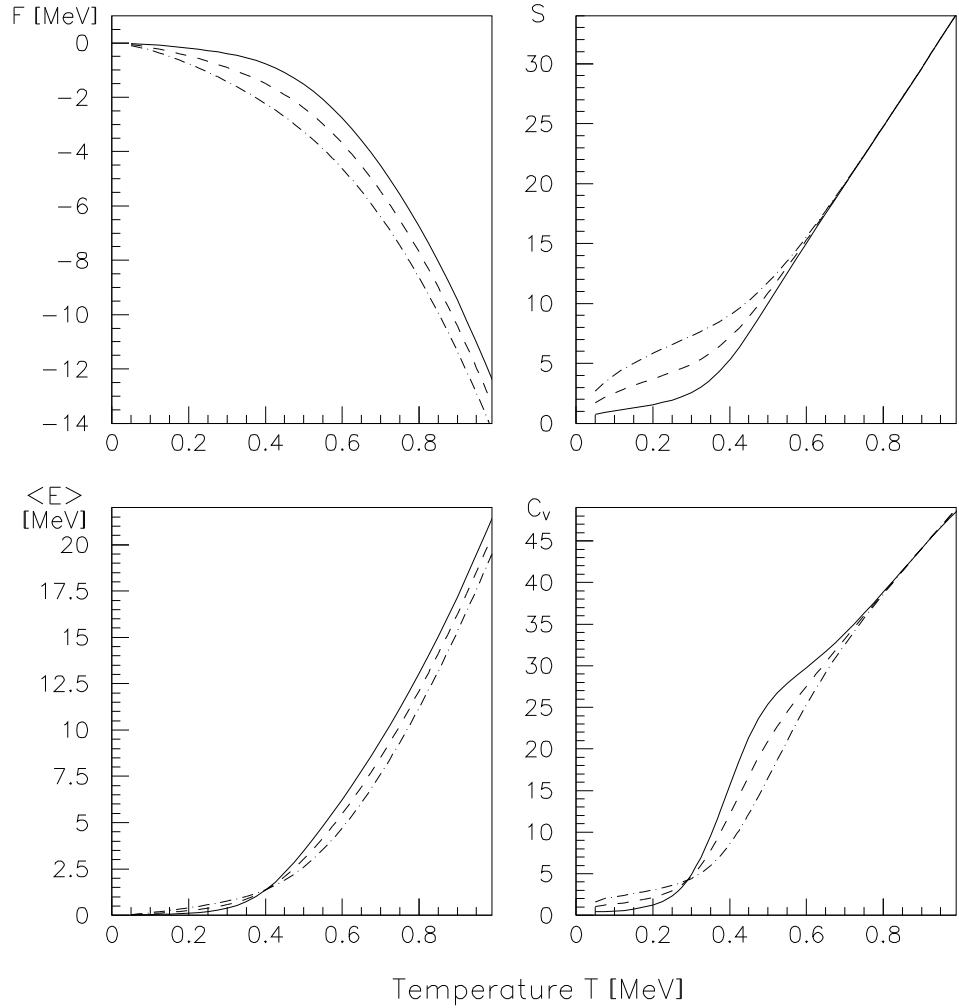} \caption{Model calculations for nuclei around $^{162}$Dy. The four panels show the free energy $F$, the entropy $S$, the thermal excitation energy $\langle E \rangle$ and the heat capacity $C_V$ as functions of temperature $T$. The same parameter set (Table I, $^{162}$Dy) is used for even-even (solid lines), odd (dashed lines) and odd-odd systems (dashed-dotted lines).} \label{fig:sfecdemo} 
\end{figure}

Figure~\ref{fig:sfecdemo} shows the Helmholtz free energy $F$, the average excitation energy $\langle E \rangle$, the entropy $S$ and the heat capacity $C_V$ as functions of temperature. The model parameters are taken from $^{162}$Dy (see Table I) as a typical set applicable for rare earth nuclei. In order to calculate the thermodynamical quantities up to $T \sim 1$ MeV, at least 10 broken nucleon pairs have to be incorporated.  
The free energy $F$ and the average excitation energy $\langle E \rangle$ behave smoothly as functions of temperature. Around $T \sim 0.65$ MeV the nuclei are excited to energies comparable to their respective neutron binding energies. The even-even, odd\footnote{We use the abbreviation {\em odd} for odd-even and even-odd systems, since these systems are equivalent in our model.} and odd-odd systems have different excitation energies at one and the same temperature, where the even-even system requires the highest $\langle E \rangle$ value.  

The entropy $S$ and heat capacity $C_V$ represent first and second derivatives of $F$, and are thus more sensible to thermal changes. For the lowest temperatures the entropy difference is $\sim 2$ between the three mass systems. However, the entropy curves coincide for $T > 0.6$ MeV, displaying almost identical behavior. It is interesting to test if our model reproduces a Fermi gas with the absence of pairing at these temperatures\footnote{A simplified Fermi gas has $S \sim 2aT + {\rm const.}$}. The level density parameter $a$ for a Fermi gas with a level spacing $\epsilon$ is given by \cite{Bohr2} 
\begin{equation}
 a=\frac{\pi^2}{6}(g_p + g_n) \sim \frac{\pi^2}{3}g=\frac{\pi^2}{3 \epsilon}. \label{eq:gpgn} 
\end{equation} 
Here, the single-particle level-density parameters for protons and neutrons ($g_p$ and $g_n$) are assumed to be approximately equal. Inserting $\epsilon=0.13$ MeV for $^{162}$Dy, we obtain $a=25$ MeV$^{-1}$. This is in exact agreement with the slope ($a=\partial S/2\partial T$) of the entropy curves for $T > 0.6$ MeV.  

The lower right panel of Fig.~\ref{fig:sfecdemo} shows the typical S-shape of the heat capacity. As this shape is an important fingerprint for pairing transitions in nuclei \cite{Schi1,Melb1,Agrawal,Liu}, we will in the following focus on its origin.  

\begin{figure} 
\includegraphics[totalheight=17.5cm,angle=0,bb=0 80 350 730]{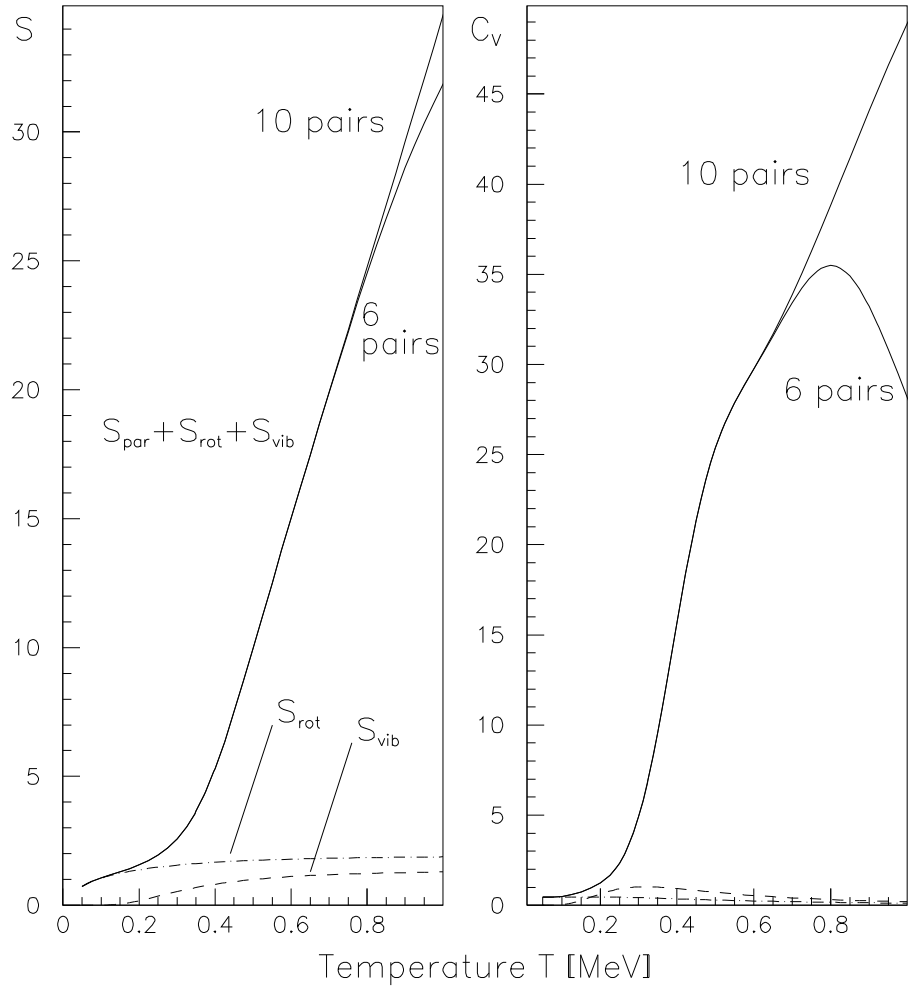} 
\caption{Entropy and heat capacity for $^{162}$Dy. The effects of the various degrees of freedom are displayed. The abrupt change when reducing the number of nucleon pairs from 10 to 6 pairs is evident.} \label{fig:scdemo} 
\end{figure}

The contribution to $C_V$ from collective excitations is negligible, and has no influence on the S shape. This is shown in Fig.~\ref{fig:scdemo} for the even-even system, where the components of particle, rotational, and vibrational degrees of freedom are displayed. The collective contribution to the entropy $S$ is small and fairly constant with increasing temperature. However, one should note that $S_{\rm rot}$ is in fact the main component at the lowest temperatures with $T < 0.3$ MeV.   

For the total entropy and heat capacity, also the effect of the number of pairs is shown in Fig.~\ref{fig:scdemo}. The $C_V$ curves are identical for 6 and 10 pairs up to $T\sim 0.7$ MeV, but from there on the 6 pair system is exhausted and not able to absorb energy at the same rate as when more Cooper pairs are present. The figure shows that the S shape can be evaluated rather accurately without taking very many pairs into account. On the other hand, if too few pairs are included, one may easily misinterpret the shape of the $C_V$ curve. For the $^{162}$Dy mass region, 6 Cooper pairs are sufficient to determine the S shape.  

\begin{figure} 
\includegraphics[totalheight=17.5cm,angle=0,bb=0 80 350 730]{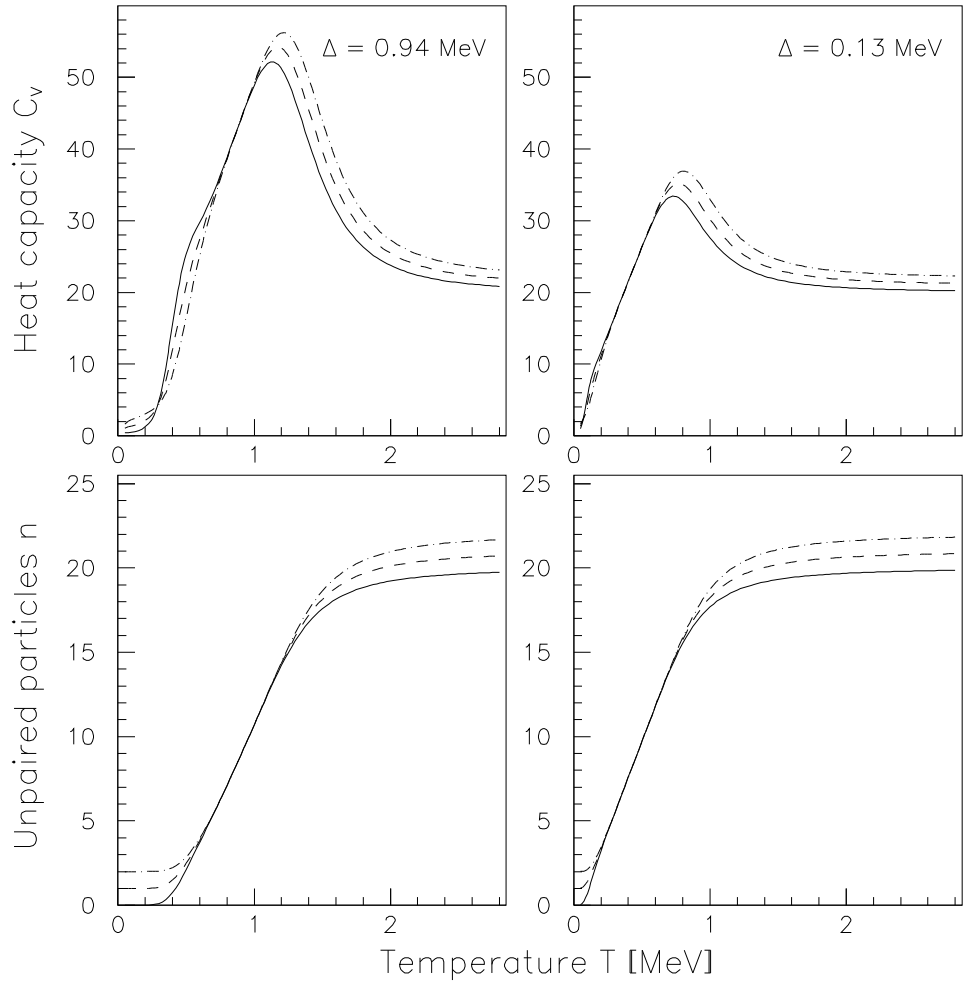} 
\caption{Heat capacity and number of unpaired nucleons for nuclei around $^{162}$Dy. In the right panels the pairing gap is reduced to $\Delta=\epsilon=0.13$ MeV. The calculations are shown for even-even (solid lines), odd (dashed lines) and odd-odd systems (dashed-dotted lines).} \label{fig:cn01} 
\end{figure}

Figure \ref{fig:cn01} shows the heat capacity (upper panels) and the corresponding number of unpaired particles (lower panels) for 10 pairs of nucleons. In the left panels a realistic pairing gap parameter of $\Delta = 0.94$ MeV has been chosen. We identify three temperature regions of interest: (i) The $T \sim 0.2 - 0.8$ MeV region with different S shapes for even-even, odd and odd-odd-systems, (ii) the Fermi gas regime with $T \sim 0.8 - 1.0$ MeV, exhibiting the same linear heat capacity for the three systems and (iii) the dramatical change for $T > 1$ MeV due to the final number of particles in the systems.

The pronounced $C_V$ maximum at $T\sim 1.0 - 1.2$ MeV is due the low number of available Cooper pairs; the lower left panel of Fig.~\ref{fig:cn01} shows that $\sim 15$ nucleons are already unpaired at this temperature. As $T \rightarrow \infty$, we find $C_V \rightarrow $ 20, 21 and 22 for the even-even, odd and odd-odd systems, respectively. This corresponds to the situation where all pairs are broken, giving a $C_V$-value equal to the total number of nucleons in the system\footnote{For the respective systems, we have 10 Cooper pairs plus 0, 1 or 2 particles.}. Since a system with one particle in an infinite harmonic oscillator gives $C_V \rightarrow 1$ for $T\rightarrow \infty$ and the effective Pauli blocking between the nucleons is negligible, we obtain $C_V=n$.

In the right panels of Fig.~\ref{fig:cn01}, the pairing gap is reduced to $\Delta= \epsilon = 0.13$ MeV. Now the S shape of the $C_V$ curve can be seen to vanish in the $\Delta \rightarrow 0$ limit, telling that the S shape is connected with the pairing strength. Furthermore, we see that the heat capacity reaches a maximum level of $C_V \sim 35$ at lower temperature $T\sim 0.7$ MeV. At this point the remaining number of particles bound in Cooper pairs is low, and the depairing mechanism again looses its capability to create more heat capacity. It might be surprising though that the number of depaired particles does not increase much faster with temperature. The answer to this is that in our model, not only the energy $2\Delta$ is required to break up a pair, but also the unpaired particles have to be placed into the single-particle level scheme at some finite energy $E$, since the Pauli blocking prevents them to populate the lowest levels in the single-particle level scheme, which are all occupied by other unpaired nucleons.

\begin{figure} 
\includegraphics[totalheight=17.5cm,angle=0,bb=0 80 350 730]{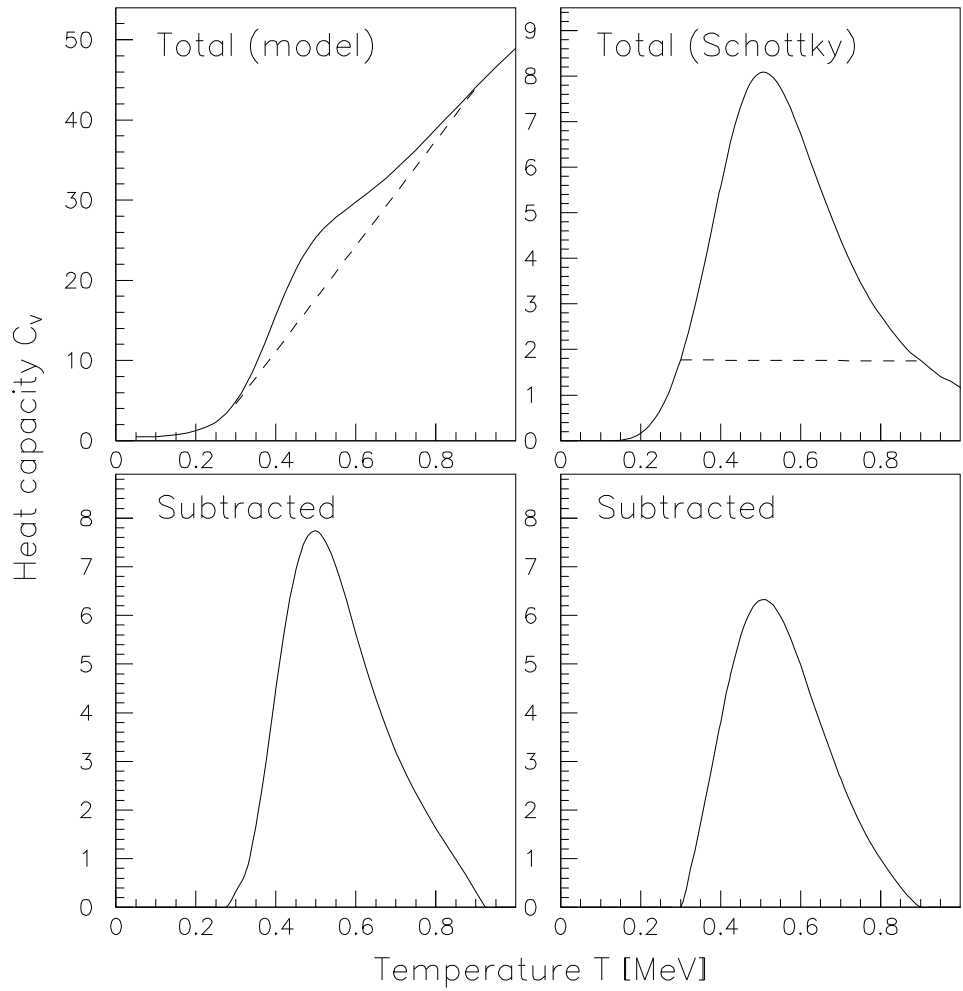} 
\caption{Heat capacity for $^{162}$Dy. The peaks in the lower panels are obtained by subtracting a linear background (dashed lines). In the right panels we have assumed a Schottky-like partition function, see text.} \label{fig:pairexemp}
\end{figure}

In Fig.~\ref{fig:pairexemp} an interpretation of the S shape is given for the even-even system. In the left panel, a "background" is subtracted from $C_V$, using a straight line through 0.3 and 0.9 MeV of temperature. This line is intended to mimic the underlying heat capacity originating from a Fermi gas type of system. However, the gas properties are strongly connected to the depairing process, and the subtracted peak in the lower left panel should be taken with care.  

The resulting peak shows some resemblance with the Schottky anomaly, which describes a particle placed in a two level system \cite{Zemansky}. However, the maximum heat capacity of such a model is $C_V \sim 0.45$ at a temperature of 40\% of the energy gap between the levels. For atomic nuclei, we may define a similar, but extended Schottky model: Either no pairs are broken at excitation energy $E=0$, or one pair is broken at $E=2\Delta$, or two pairs are broken at $E=4\Delta$ and so on. This picture is an extreme simplification, and we include here only zero, one and two pairs. Thus, our Schottky-like partition function reads 
\begin{equation}
 Z_{\rm Sch}=1+W_{2\Delta}e^{-2\Delta/T}+W_{4\Delta}e^{-4\Delta/T}. \label{eq:schottky} 
\end{equation} 
The multiplicities in front of the Boltzmann factors represent the number of levels at $E=2\Delta$ and $E=4\Delta$ and are estimated from experimental data using $W=\rho \cdot \delta E$, where $\rho$ is the level density and $\delta E$ is the energy window considered.  

In the right panels of Fig.~\ref{fig:pairexemp} $C_V$ is evaluated by the use of Eq.~(\ref{eq:schottky}) for $^{162}$Dy. The $\rho$ values are estimated from Table I of Ref.~\cite{Gutt2}. With an energy window of $\delta E=0.2$ MeV, we obtain $W_{2\Delta}=22$ and $W_{4\Delta}=556$; numbers which are also consistent with the numbers of seniority $\cal S =$ 2 and 4 states in the model of Ref.~\cite{Gutt1}. The high multiplicities give a strength, width, and position of the peak that compare qualitatively well with the peak in the lower left panel. From this comparison, we interpret the local maximum of the S-shaped $C_V$ curve at $T\sim 0.5$ MeV as the point at which the depairing process is at the strongest. Hence, we define this point as the critical temperature $T_c$ for the pairing transition.  

To close this section, we comment on the weakening of the S shape when going from even-even to odd and odd-odd systems; an effect which is clearly seen in the lower right panel of Fig.~\ref{fig:sfecdemo}. The reason for the weakening can be understood from the entropy plot displayed in the right upper panel of Fig.~\ref{fig:sfecdemo}. There, we saw that the entropy of each valence nucleon\footnote{Interpreted as the entropy gap between the various S-curves \cite{Gutt1}.} is reduced from a value of $S\sim 2$ at $T\sim 0.3$ MeV to zero at $T\sim 0.6$ MeV. Since the corresponding heat capacity relates to $T\partial S/ \partial T$, a negative contribution to $C_V$ appears and a quenching of the amplitude of the S-curve is apparent.

\section{Comparison with data}  

\begin{figure} 
\includegraphics[totalheight=17.5cm,angle=0,bb=0 80 350 730]{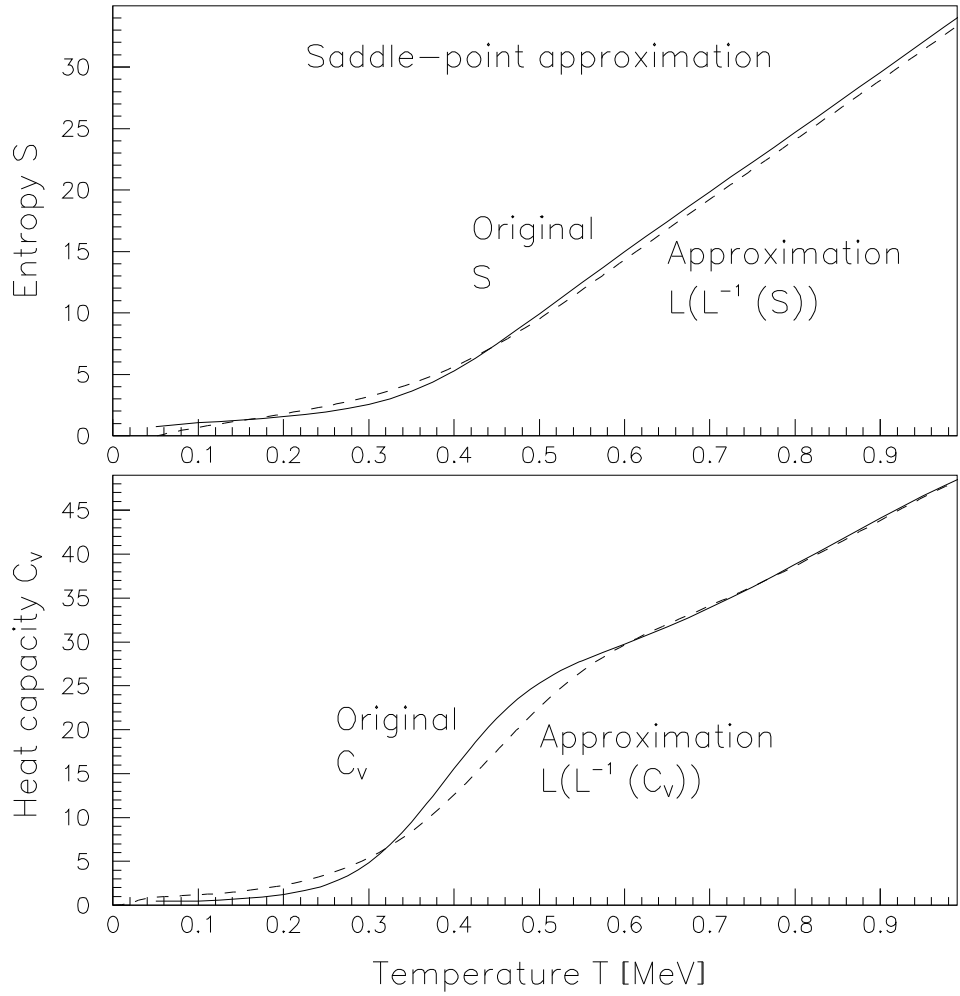} 
\caption{Test on the saddle-point approximation. The solid lines are entropy and heat capacity obtained from our model for $^{162}$Dy. The dashed lines are calculated by making an inverse Laplace transformation with the saddle-point approximation and then back again with a Laplace transformation, see text.} \label{fig:laplace}
\end{figure}

Our model is described within the canonical ensemble, while experimental data refer to the microcanonical ensemble. However, there are two ways to compare our model with experiments. With a known experimental level density $\rho$, the partition function can be constructed from 
\begin{equation}
Z(T)=\sum _i \delta E_i \rho(E_i) e^{-E_i/T} \label{eq:laplace},
\end{equation}
where $E_i$ is the excitation energy and $\delta E_i$ are the energy bins used. From this partition function all thermodynamical quantities can be deduced, see e.g.~Eqs.~(\ref{eq:FT}-\ref{eq:CV}). The drawback of this method, is that the level density function has to be known up to high excitation energies, typically $E\sim 40$ MeV. The other way is to evaluate the microcanonical level density from our canonical model. This can be performed by an inverse Laplace transformation of $Z$. Using the saddle-point approximation (Fawler-Darwin method), we obtain\cite{Koonin} 
\begin{equation}
\rho(\langle E\rangle)=\frac{e^S}{T\sqrt{2 \pi C}}. \label{eq:rho} 
\end{equation}
The level density can then be compared with certain experimental values, and complete knowledge on the level density up to high excitation energies is not necessary. One drawback using the saddle-point approximation is that the excitation energy $\langle E\rangle$ is a thermal average with a large standard variation of $\sigma_E=T\sqrt{C}\sim T\sqrt{2aT}$, with $a$ being the level-density parameter. A second drawback is the approximation itself, which we have tested by a  "forward-backward" Laplace transformation. In Fig.~\ref{fig:laplace} we compare the original entropy and heat capacity, with the ones obtained by using Eq.~(\ref{eq:rho}) to obtain $\rho$, and then using Eq.~(\ref{eq:laplace}) to obtain a new $Z$ and its corresponding $S$ and $C_V$. The comparison reveals a general smoothing for temperatures $T < 0.5 - 0.6$ MeV. The entropy is seen to be reproduced rather well, however, the heat capacity is more sensitive to the approximation.

The model presented in Sec.~II rests on the assumption that the single-particle level scheme can be approximated by equidistant levels. This assumption is never fulfilled in atomic nuclei, but within the canonical ensemble, the various deduced thermodynamical quantities are strongly smoothed with respect to excitation energy. Therefore, even with some non-uniformity, the model might still give realistic results. Probably, heavy and strongly deformed nuclei are the best candidates for our model.

For test cases, we have chosen midshell nuclei around $^{58}$Fe, $^{106}$Pd, $^{162}$Dy and $^{234}$U. All four mass regions reveal a rather uniform Nilsson single-particle energy distribution without large energy gaps. The pairing gap $\Delta$ and rotational parameter $A_{\rm rot}$ are calculated according to Eqs.~(\ref{eq:gap}) and (\ref{eq:arot}). The vibrational energy quantum $\hbar \omega_{\rm vib}$ is taken as the energy at which the first vibrational states appear in the experimental level schemes of the respective mass region \cite{Firestone}. The last parameter needed is the level-gap parameter $\epsilon$, which is expected to have a value in between the single-particle spacing $\epsilon_{\rm sp}$ and the BCS quasiparticle spacing $\epsilon_{\rm qp}=\sqrt{(\epsilon_{\rm sp}-\lambda)^2+\Delta^2}-\Delta$, where $\lambda$ is the Fermi level.

In this work, $\epsilon$ is chosen as a free parameter, determined from a fit to known experimental level densities. For each nucleus we adopt two level density anchor points, as deduced in Ref.~\cite{Gutt2}. The lower anchor point is based on the counting of known discrete levels. This method is rather accurate, except for the odd-odd nuclei, where the number of levels might be several hundred per MeV, and thus difficult to measure. The other anchor point is based on average neutron-resonance spacing data at the neutron-binding energy.  

\begin{figure} 
\includegraphics[totalheight=17.5cm,angle=0,bb=0 80 350 730]{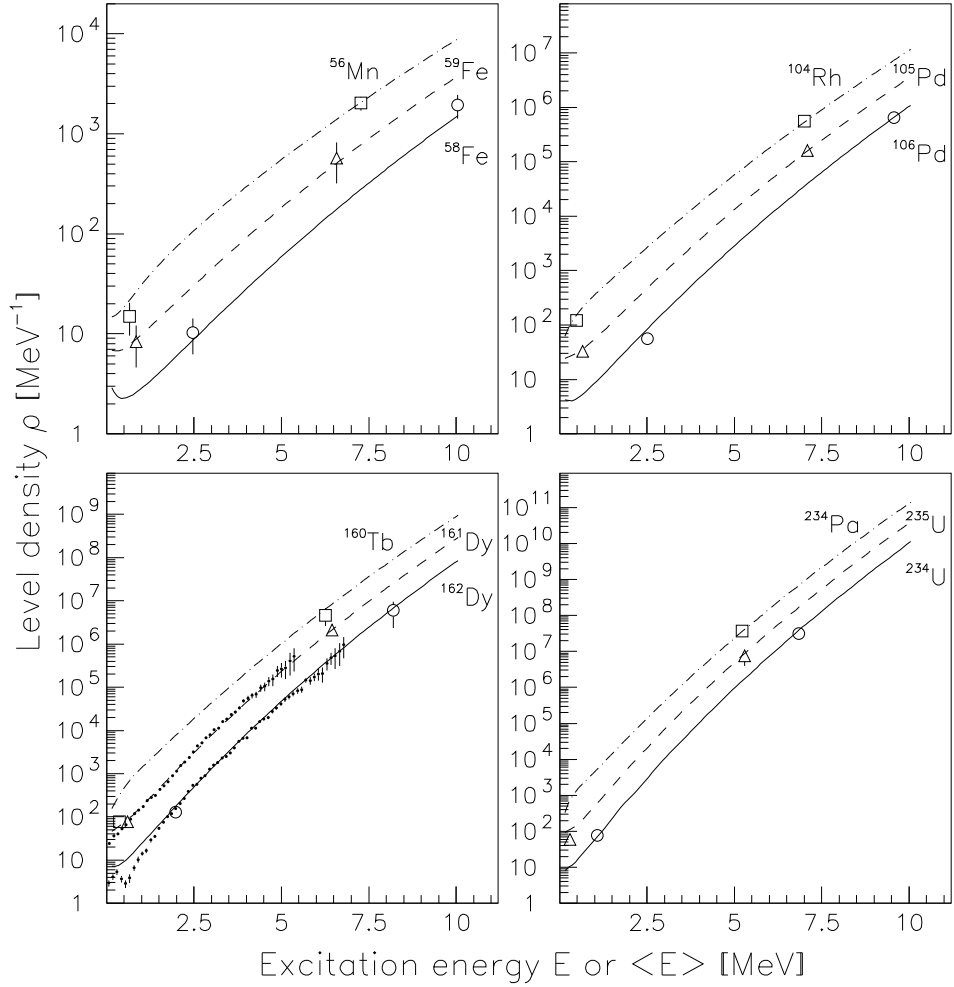} 
\caption{Calculated level densities of even-even (solid line), odd (dashed line) and odd-odd (dashed-dotted line) nuclei around $^{58}$Fe, $^{106}$Pd, $^{162}$Dy and $^{234}$U as function of average excitation energy $\langle E \rangle$. The open circles, squares and triangles are experimental level-density anchor points [11] extracted at certain excitation energies $E$. The solid circles for $^{161,162}$Dy are experimental data points from Ref.~[4].} 
\label{fig:rhofepddyu} 
\end{figure}

Figure \ref{fig:rhofepddyu} shows anchor points and level densities\footnote{The experimental data are taken at a given excitation energies $E$, while the calculations give $\rho$ as function of average excitation energy $\langle E\rangle$.} calculated using Eq.~(\ref{eq:rho}). The parameters are listed in Table I, where $\epsilon$ is adjusted to obtain the approximate slope of $\rho$ in the log-plot. The agreement with the anchor points is good, and also the odd and odd-odd systems fall nicely into the systematics. For the $^{161,162}$Dy isotopes, the experimental level densities from Ref.~\cite{Schi1} are shown as well. The calculations reveal good agreement with experiment as function of excitation energy.

\begin{figure} 
\includegraphics[totalheight=17.5cm,angle=0,bb=0 80 350 730]{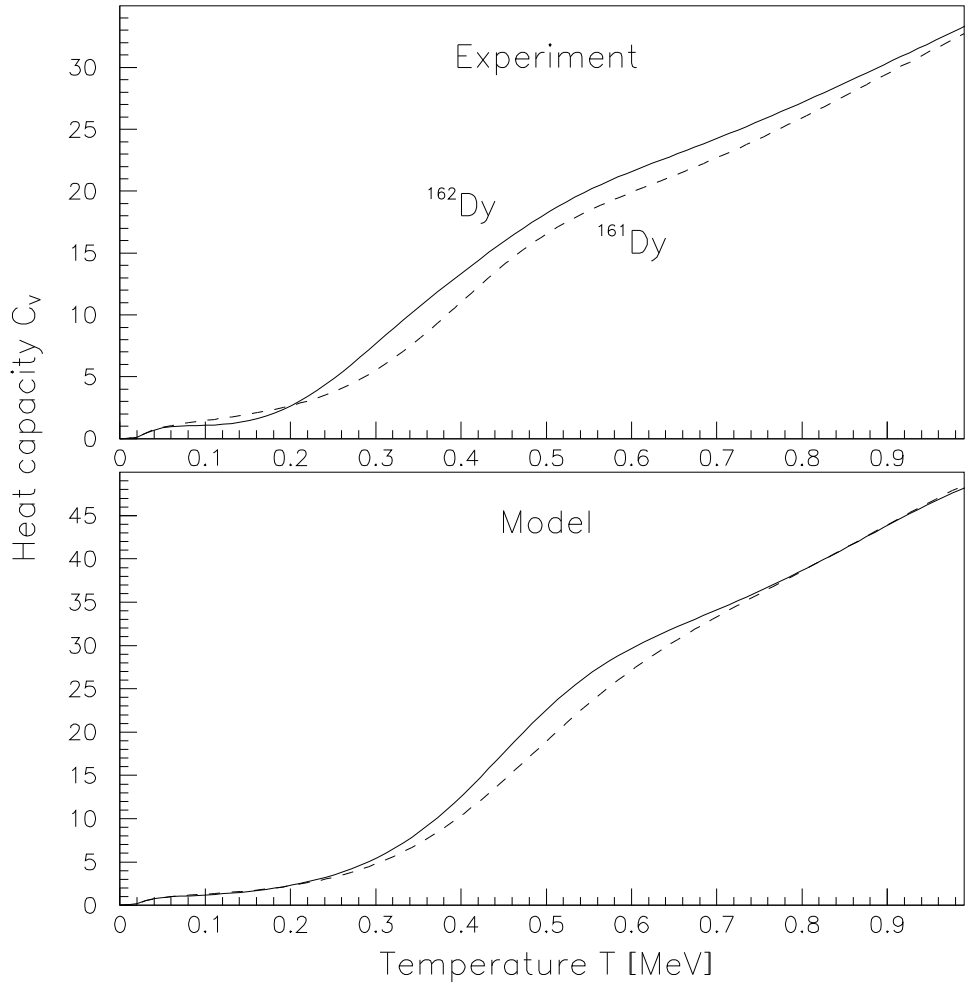} 
\caption{Comparison between semiexperimental (upper panel) and theoretical (lower panel) heat capacity for $^{161,162}$Dy. The semiexperimental values are taken from Ref.~[4].} \label{fig:cvteoexp}
\end{figure}

With the established parameters, we may now calculate the heat capacity $C_V$ from our model and compare with experiments. Due to the imperfection of the saddle-point approximation, we take $\rho(\langle E \rangle)$ from the saddle-point approximation (which is tuned to experiments in Fig.~\ref{fig:rhofepddyu}), and generate $C_V$ from the corresponding $Z$ function of Eq.~(\ref{eq:laplace}). In the lower part of Fig.~\ref{fig:cvteoexp} the theoretical $C_V$ curves are shown for $^{161,162}$Dy. The experimental $C_V$ curves, shown in the upper part, are based on  experimental level density data from Ref.~\cite{Schi1}. Since the construction of $Z$ requires data to much higher energies than experimentally known ($E \sim 7-8$ MeV), the level density has to be extrapolated. Here, we have used the parametrization of Egidy et al.~\cite{Egidy}, in accordance with our recent work \cite{Schi1}. The effective level density parameter ($a \sim 17.5$ MeV$^{-1}$) has a major impact on the $C_V$ curve for $T> 0.5$ MeV. Figure \ref{fig:cvteoexp} shows rather good agreement between experiment and model for $T < 0.5 - 0.6$ MeV. In particular the odd-even mass entropy difference is well reproduced. At higher temperatures the comparison is poor due to the arbitrary extrapolation of the experimental level densities.

\begin{figure} 
\includegraphics[totalheight=17.5cm,angle=0,bb=0 80 350 730]{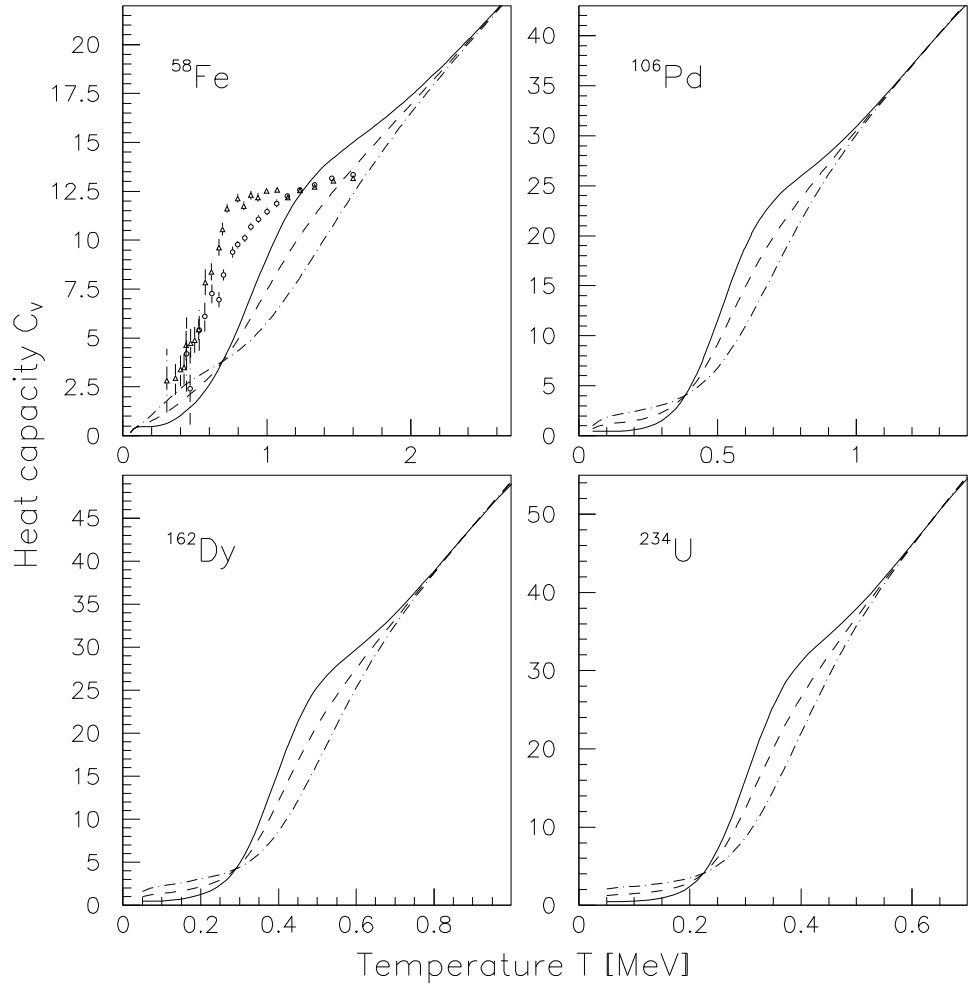} 
\caption{Calculated heat capacities for even-even (solid lines), odd (dashed lines) and odd-odd (dashed-dotted lines) nuclei around $^{58}$Fe, $^{106}$Pd, $^{162}$Dy and $^{234}$U. In addition, the heat capacities from SMMC simulations (scanned from Fig.~4 of Ref.~[9]) are displayed as open triangles ($^{58}$Fe) and circles ($^{59}$Fe).} \label{fig:fourcvt}
\end{figure}

In Fig.~\ref{fig:fourcvt} the heat capacity $C_V$ from our model, using the parameters of Table I, is displayed for the four mass regions. The curves look similar; the differences are mainly the change in the scaling of the $C_V$ and $T$ axes for the various mass regions. The figure also includes the calculations on $^{58,59}$Fe performed by Liu and Alhassad \cite{Liu}. These data points show some discrepancies with the work of Rombouts {\em et al.}~\cite{Rombouts}. However, both SMMC calculations obtain a critical temperature around 0.7 MeV, while we obtain a value around 1.2 MeV. Even when changing freely the number of particles, $\epsilon$ and $\Delta$ in our model, we are not able to reproduce the SMMC results. In particular, the feature that $C_V$ becomes constant \cite{Liu}, or even drops \cite{Rombouts} above $T\sim 0.7$ MeV is surprising. This effect might originate from shell gaps (which are not included in our model) or too few particles and/or orbitals considered in the SMMC calculations. It would be interesting to test if the SMMC calculations are capable of reproducing the anchor points for Fe in Fig.~\ref{fig:rhofepddyu}. Doing so, one should remember that our anchor points represent densities of levels and not states, which are usually employed in SMMC calculations.   

\section{Conclusions}  
A simple model for hot nuclei has been outlined. The main properties of the model are determined by the pairing-gap parameter $\Delta$ and the energy-gap parameter $\epsilon$ associated with an infinite single-particle level scheme for protons and neutrons. We have demonstrated how various thermodynamical quantities can be extracted.

The model properties have been discussed with emphasize on the S shape of the heat capacity. This shape is strongly related to the bunch of newly created levels from the pair-breaking process at $\sim 2 \Delta$ and $\sim 4 \Delta$ of excitation energy.   

The model calculations are compared with experimental data from the $A \sim$ 58, 106, 162 and 234 mass regions. Using the saddle-point approximation with only one free parameter, the experimental level densities of even-even, odd and odd-odd systems are reproduced.  

The critical temperature of the pair transition for $^{161,162}$Dy was calculated to be $T_c \sim 0.5$ MeV, in agreement with experiments. In the $^{58,59}$Fe region we calculate $T_c \sim 1.2$ MeV, which is significantly higher than the value of $\sim 0.7$ MeV obtained in Shell Model Monte Carlo simulations. Thus, further theoretical and experimental efforts are needed to understand the thermodynamics of these hot iron isotopes.  

 We wish to acknowledge the support from the Norwegian Research Council (NFR).

\end{multicols}   

\begin{table} Table I: Model parameters: $\Delta$ and $A_{\rm rig}$ are taken from Eqs.~(\ref{eq:gap}) and (\ref{eq:arot}), $\hbar \omega_{\rm vib}$ from systematics, and $\epsilon$ is tuned to fit the data of Fig.~\ref{fig:rhofepddyu}.\\ \begin{tabular}{l|c|c|c|c} 
Nucleus & $\Delta$ & $A_{\rm rig}$&\ $\hbar \omega_{\rm vib}$  & $\epsilon$\\ 
        &  (MeV)   &   (keV)      &    (MeV)      &     (MeV)      \\ \hline &&&&\\ $^{58}$Fe  & 1.58 & 42.0          & 2.0           &     0.80     \\ 
$^{106}$Pd & 1.17 & 15.4          & 1.4           &     0.21     \\
 $^{162}$Dy & 0.94 &  7.6          & 0.9           &     0.13     \\ 
$^{234}$U  & 0.78 &  4.1          & 0.8           &     0.08     \\ \end{tabular}
\end{table}

\end{document}